\documentclass[aps,prl,twocolumn,superscriptaddress,showpacs]{revtex4-2}

\usepackage{amsmath,amssymb}
\usepackage{bm}
\usepackage{graphicx}
\usepackage{mathtools}
\usepackage{hyperref}
\usepackage{xcolor}%

\newcommand{\E}{\mathbb{E}}
\newcommand{\N}{\mathcal{N}}
\newcommand{\du}{\delta u}
\newcommand{\M}{\bm{M}}
\newcommand{\jbar}{\bar{j}}
\newcommand{\gj}{\gamma_j}
\newcommand{\tauj}{\tau_j}

\begin{document}

\title{Inferring Multi-Scale Cascade Structure from a Single Lagrangian Trajectory via Hidden Markov Switching}

\author{Ziang Yang}
\affiliation{Department of Civil and Environmental Engineering, University of Pittsburgh, Pittsburgh, PA 15261, USA}

\author{Lei Fang}
\email[Corresponding author: ]{lei.fang@pitt.edu}
\affiliation{Department of Civil and Environmental Engineering, University of Pittsburgh, Pittsburgh, PA 15261, USA}
\affiliation{Department of Mechanical Engineering and Materials Science, University of Pittsburgh, Pittsburgh, PA 15261, USA}
\affiliation{Department of Bioengineering, University of Pittsburgh, Pittsburgh, PA 15261, USA}
\date{\today}

\begin{abstract}
We show that a single Lagrangian trajectory encodes not only temporal but also spatial cascade structure—recoverable without access to a spatially resolved Eulerian velocity field. We report a Markov-Switching Multifractal (MSM) framework that infers hidden multi-scale cascade states directly from single-particle trajectories.
Kolmogorov scaling fixes the model's switching rates, leaving a single calibrated intermittency parameter; exact Bayesian filtering then returns the posterior of the layer-resolved cascade state along the trajectory.
When we apply the MSM framework independently to each trajectory of a particle pair, we find that the inferred states carry genuine spatial information: consistent with Richardson’s locality picture, the Kolmogorov $r^{1/3}$ scaling is preserved when fine-scale layers are randomized, but collapses once the randomization crosses the layer set by the pair separation. Single trajectories thus encode the spatial organization of the cascade, opening a route to cascade diagnostics for turbulent flows accessible primarily through sparse or individual Lagrangian trajectories—from particle-tracking velocimetry to atmospheric and oceanographic tracking—where the instantaneous Eulerian field is out of reach.
\end{abstract}

\maketitle


Turbulent cascades are intrinsically spatial: energy is organized and transferred across a hierarchy of eddy scales \cite{Frisch1995, Sreenivasan1997}. Yet many observations of turbulence are inherently Lagrangian, consisting only of the kinematic history along individual particle trajectories. Whether an individual trajectory retains information about the surrounding spatial hierarchy of the cascade remains unknown.

The spatial hierarchy of the cascade manifests through strongly intermittent, scale-dependent fluctuations: velocity increments and dissipation become increasingly bursty toward small scales, a multiscale statistical structure captured by multifractal descriptions of turbulence.
This structure has been characterized in many ways---through the scale-to-scale multiplier distributions of the coarse-grained field~\cite{Chhabra1992, Sreenivasan1995}, its multifractal spectrum via wavelets~\cite{Farge1992, Muzy1991}, and, treated as a Markov process in scale, the stochastic equation governing the increments across scales~\cite{Friedrich1997, Renner2001}. These characterizations share one prerequisite: access to the spatially resolved field, from which structure can be coarse-grained across scales.


This raises a key question about the cascade itself: how much information about its spatial, multiscale structure is carried by a single Lagrangian trajectory?
A trajectory samples the flow along one path in space and time, yet the cascade is an inherently spatial, multiscale object---so it is far from obvious whether one trajectory contains the local cascade state at all. 

This question is not only fundamental but also important in practice, because a large portion of turbulence observed is Lagrangian: laboratory particle-tracking velocimetry~\cite{LaPorta2001, Toschi2009}, atmospheric balloons \cite{MorelLarcheveque1974,Lacorata2004}, and oceanographic drifters \cite{RichardsonStommel1948,Lumpkin2017} all report the flow along individual trajectories, with little or no access to the surrounding instantaneous field. The extent to which a single such trajectory encodes the state of the multiscale cascade sets a basic limit on what these observations can reveal about the underlying turbulence.

In this Letter, we answer this question by applying a turbulent-physics-constrained Markov-Switching Multifractal (MSM) model~\cite{Calvet2004} to Lagrangian trajectories in turbulent flows. The layer switching rates are not fitted but determined by Kolmogorov scaling, reducing the model to a single calibrated parameter. Because the latent state is low-dimensional and discrete, Bayesian filtering is exact. Thus, the MSM model returns the causal posterior of the layer-resolved cascade state along a single trajectory. 
As consistency checks, the inferred model reproduces the anomalous Lagrangian structure-function exponents and the Reynolds-number-dependent, non-Gaussian acceleration statistics observed over $Re_\lambda = 200\text{--}815$~\cite{Xu2006, Mordant2004}. In order to understand the extent to which a single trajectory encodes the spatial organization of the cascade, we apply the MSM model independently to each member of particle pairs. The states inferred from purely temporal single-trajectory data encode the spatial scale set by the pair separation: reconstructing the trajectories recovers Kolmogorov $r^{1/3}$ scaling, and this scaling is insensitive to randomizing the fast, fine-scale states---until the truncation reaches the layer set by the separation, where it breaks down sharply. Single Lagrangian trajectories thus natively encode the spatial cascade generator, opening a route to field-free cascade diagnostics for particle-tracking velocimetry and subgrid-scale modeling.


\emph{Model.}---Let $\bm{u}(t)$ be the velocity of a Lagrangian tracer sampled at interval $dt$.  The observed increment magnitude $\du(t) = |\bm{u}(t+dt) - \bm{u}(t)|$, conditional on the latent state $\M_t$, is half-normal with scale $\sigma\sqrt{g(\M_t)}$,
\begin{equation}
  \du(t) = \sigma \sqrt{g(\M_t)}\;|\varepsilon_t|,
  \qquad
  \varepsilon_t \overset{\mathrm{iid}}{\sim} \N(0,1),
  \label{eq:msm}
\end{equation}
where $g(\M_t) = \prod_{j=1}^{\jbar} M_{j,t}$ and the multipliers $M_{j,t} \in \{m_0,\, 2-m_0\}$, $m_0 \in (1,2)$, are independent two-state Markov chains with $\E[M_{j,t}] = 1$, which under the assumption of layer independence gives $\E[\du^2] = \sigma^2$. The increment scale $\sigma = \du_{\mathrm{rms}}$ is measured directly from the trajectory data where $\du_{\mathrm{rms}}$ is the root-mean-square velocity increment along trajectories.
The only calibrated intermittency parameter is $m_0$.  Each layer $j$ evolves by the symmetric transition matrix
\begin{equation}
  A_j =
  \begin{pmatrix}
    1 - \gj/2 & \gj/2 \\
    \gj/2     & 1 - \gj/2
  \end{pmatrix},
  \label{eq:Pi}
\end{equation}
i.e.\ with probability $\gj$ the state is redrawn from the symmetric two-point distribution (falling back on the same value half the time). The switching rate is not fitted but prescribed by the physics: layer $j$ is identified with eddies of Lagrangian correlation time
\begin{equation}
  \tauj = \tau_\eta\, b^{\,\jbar - j},
  \qquad b = 2^{2/9},
  \label{eq:tauk}
\end{equation}
where $b$ is the layer-to-layer ratio of \emph{time} scales, obtained from the factor-of-two spatial hierarchy through the Lagrangian scaling $\tau(\ell) \sim \varepsilon^{-1/3}\ell^{2/3}$ (see Supplemental Material (SM)), and $\gj = 1 - e^{-dt/\tauj}$ is fixed by requiring the exact autocorrelation at lag $\Delta t$
\begin{equation}
  \mathrm{Corr}\bigl[M_{j,t+\Delta t},\, M_{j,t}\bigr]
  = e^{-\Delta t/\tauj}.
  \label{eq:corr}
\end{equation}
The number of layers is fixed by the inertial-range extent: with the spatial hierarchy $\ell_j = \eta\,\lambda^{\jbar-j}$ of ratio $\lambda = 2^{1/3}$ per layer (a factor of two per three layers), the condition $\ell_1 \simeq L$ gives $\jbar = \lfloor 3\log_2 (L/\eta) + 1\rfloor$, where $L$ is the integral length scale and $\eta$ is the Kolmogorov length scale; the time-scale ratio in Eq.~\ref{eq:tauk} then follows as $b = \lambda^{2/3} = 2^{2/9}$.

In this Letter, groups of three consecutive layers are aggregated into $\bar{k} = \lfloor\jbar/3\rfloor$ blocks: within block $k$ the three binary multipliers enter the observation likelihood only through their product $\mathcal{M}_{k,t}$, which takes four values indexed by the number of layers in the high state. In this way, the length scale ratio between $k$ and $k+1$ is 2. The aggregation preserves exactly the product $g(\M_t)$ entering the likelihood and the Markov dynamics of the block products (whose $4\times4$ transition matrices follow from marginalizing the micro-chains). The states of individual layers within a block are not resolved. The posterior over the $4^{\bar{k}}$ block macro-states is then computed exactly by the recursive Bayes update
\begin{equation}
  P_t \propto \bm{v}\bigl(\du(t)\bigr) \odot \bigl(P_{t-1} A\bigr),
  \label{eq:bayes}
\end{equation}
where $\bm{v}$ is the vector of observation likelihoods and $A$ the Kronecker-product transition matrix.  All state-resolved results below refer to block macro-states.

In this Letter, we use trajectories extracted from the Johns Hopkins Turbulence Database forced isotropic dataset ($1024^3$, $Re_\lambda = 433$) \citep{Li2008,Yu2012}, with $\tau_\eta$, $\eta$, and $L$ taken from the DNS.  
Hidden states of trajectories are sampled sequentially. At each time step, the previously sampled state is propagated using the transition matrix $A$, the probabilities of the possible states are updated using the current observation, and a new state is drawn from this updated distribution. Repeating this procedure produces Markov paths conditioned only on observations available up to the current time.

\begin{figure}[t]
  \centering
  \includegraphics[width=\columnwidth]{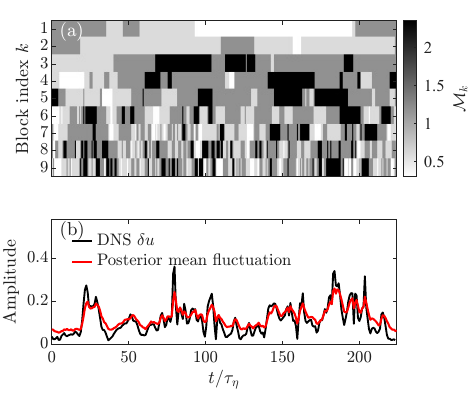}
  \caption{(a) One filtered-posterior draw of the hidden block macro-states $\mathcal{M}_k(t)$, for a single Lagrangian trajectory ($Re_\lambda=433$). Darker shading indicates larger multiplier values. (b) Posterior mean fluctuation inferred from 200 filtered-posterior samples (red), overlaid on the observed DNS velocity increment magnitude $\delta u$ (black).}
  \label{fig:states}
\end{figure}

\emph{Hidden cascade states.}---Fig.~\ref{fig:states} shows one filtered-posterior draw of the block macro-states $\mathcal{M}_k(t)$, $k = 1,\ldots,\bar{k}$, for a single trajectory in the upper panel, while the lower panel shows the posterior mean obtained from 200 independent filtered-posterior samples overlaid on the DNS increment magnitude $\du(t)$.
The effective sampling interval is $dt\approx\tau_\eta$. 
The time-scale difference imposed by Eq.~\ref{eq:corr} is evident: large-scale states persist over a long period of time while smaller-scale states fluctuate and switch increasingly quickly. The posterior mean over independent draws tracks the DNS fluctuation envelope closely, confirming that the filter extracts a stable multi-scale representation from a single trajectory.

\begin{figure}[t]
  \centering
  \includegraphics[width=\columnwidth]{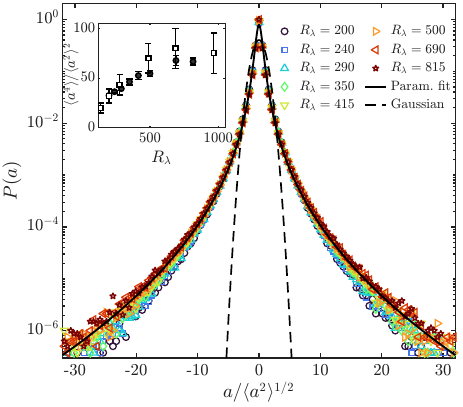}
  \caption{Normalized acceleration PDF from forward MSM simulation (no DNS input) for eight Reynolds numbers ($Re_\lambda = 200$--$815$), with the layer count of each fixed by the measured $L/\eta$ of the corresponding experiment.  Solid line: parameterized stretched-exponential fit. Dashed line: Gaussian.  Inset: acceleration flatness versus $Re_\lambda$ (mean $\pm$ standard error of the mean over 12 realizations). Filled circles denote the MSM results, while open squares denote the experiment of La Porta \textit{et al.}~\cite{LaPorta2001} with the reported error bars.}
  \label{fig:accel}
\end{figure}

\emph{Intermittency.}---As a consistency check, we first verify that the model, with its physically prescribed switching rates, reproduces Lagrangian intermittency over a range of Reynolds numbers. With the single intermittency parameter $m_0$ calibrated once by matching the forward MSM's acceleration and structure-function statistics to experiments \cite{Xu2006}, with fixed $m_0$ for all Reynolds numbers, the MSM is run forward as a generative model---no DNS data or Bayesian inference is involved.
Signed increments $\delta u_t = \sigma\sqrt{g(\M_t)}\,\varepsilon_t$ are drawn directly (the sign carried by $\varepsilon_t$) from the model. The velocity record is reconstructed by summation $u(t) = \sum_{t'\leq t}\delta u_{t'}$ over time. Structure functions at lag $\tau$ are computed from differences of the reconstructed record, and the acceleration is estimated by local quadratic fits over windows of $0.75\,\tau_\eta$, as in particle-tracking practice \citep{Voth2002}.
Across $Re_\lambda = 200$--$815$, only the layer count $\jbar$ is changed physically based on the $L/\eta$ of the corresponding experimental conditions \citep{Voth2002}; $m_0$ is not refitted per Reynolds number, and $\sigma$ drops out of all normalized statistics.  Fig.~\ref{fig:accel} shows the resulting normalized acceleration probability density function. The central portions of the distributions approximately collapse, with stretched-exponential tails extending up to $30\,\langle a^2\rangle^{1/2}$ in agreement with experiment \citep{LaPorta2001,Mordant2004}, while increasingly rare tail events produce a Reynolds-number-dependent flatness (inset), reflecting the deepening of the cascade.

\begin{figure}[t]
  \centering
  \includegraphics[width=\columnwidth]{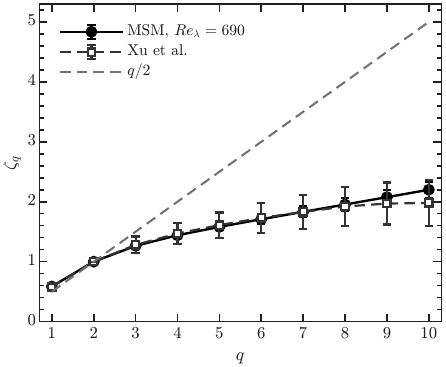}
  \caption{Relative scaling exponents $\zeta_q/\zeta_2$ of Lagrangian structure functions at $Re_\lambda = 690$. Filled circles denote MSM results (mean $\pm$ standard error of the mean over 20 realizations). Open squares denote the experiment of Xu \emph{et al.} \citep{Xu2006}.}
  \label{fig:zeta}
\end{figure}

Lagrangian structure functions
$S_q(\tau)=\langle|\delta u_S^L(\tau)|^q\rangle\sim\tau^{\zeta_q}$
scale anomalously, where $\delta u_S^L(\tau)=u(t+\tau)-u(t)$ denotes the Lagrangian velocity increment along a trajectory at lag $\tau$.
Their exponents depart from the K41 value $\zeta_q=q/2$~\cite{Xu2006}.
Extracting $\zeta_q/\zeta_2$ by extended self-similarity, the MSM exponents agree with the measurements of \citet{Xu2006} within error bars across $q = 1$--$10$ (Fig.~\ref{fig:zeta}), exhibiting the concave departure from the K41 prediction $\zeta_q = q/2$ that is the signature of multiplicative intermittency.

\emph{Spatial cascade structure revealed by a single particle.}---
We now come to the key question: since a trajectory carries mainly temporal information, can the state inferred from it nonetheless reveal the spatial organization of the cascade? 
To answer this question, we specifically ask whether the layers inferred from a single trajectory correspond to physical ``eddies" of a corresponding length scale, despite the trajectory carrying no explicit spatial information. Particle pairs provide a stringent test through two established facts. First, in our construction, each block $k$ is tied to a spatial scale $\ell_k$ by Kolmogorov scaling, and we define $k(r)$ as the index of the block corresponding to eddies of length scale $r$. 
Second, 
Kolmogorov inertial-range scaling \cite{Kolmogorov1941} states that the relative velocity obeys
\begin{equation}
  \delta u^{E}(r) \sim (\varepsilon r)^{1/3},
  \label{eq:r13}
\end{equation}
where $\delta u^{E}(r)$ denotes the Eulerian velocity difference across a spatial separation $r$, and the superscript $E$ distinguishes it from the Lagrangian velocity increment.



For each pair, we filter the two trajectories independently, using the radial velocity increment $\delta u_r^{(i)} = \delta\bigl[\bm{u}^{(i)}\!\cdot\hat{\bm{r}}\bigr]$ as the observable, since the separation rate is governed by the radial velocity difference. Each block $k$ carries a spatial correlation length $\ell_k$, with the length scales dyadically spaced from $L$ down to $\eta$.  From the posterior samples, we form the family of intermittency kernels

\begin{equation}
  w_{s,i}(t)
  =
  \left[
    \Pi^{(i)}_{1:k(r)+s}(t)\,
    \widetilde{\Pi}^{(i)}_{k(r)+s+1:\bar{k}}(t)
  \right]^{1/2},
  \label{eq:kernels}
\end{equation}
where
$\Pi^{(i)}_{a:b}(t)
= \prod_{j=a}^{b}\mathcal{M}^{(i)}_j(t)$
denotes the product over blocks $a$ through $b$, and
$\widetilde{\Pi}^{(i)}_{k(r)+s+1:\bar{k}}(t)$ denotes a stationary random
completion of the blocks not retained from the posterior, with the shifted
cutoff restricted to the available block range. The integer shift $s$ controls how many blocks beyond the physical cutoff are retained; $s=0$ truncates exactly at the current separation, and the all-block kernel $a_i=\sqrt{\Pi_{1:\bar{k}}^{(i)}}$ serves as reference.
 The absolute pair difference is integrated along each sample path, $X_s(t) = \sum_{t'\leq t} \bigl|w_{s,2}(t') - w_{s,1}(t')\bigr|\,dt$, and averaged over all sample times at which the pair separation lies within a logarithmically spaced interval $[r,\,r+\mathrm{d}r)$ centered on $r$,
\begin{equation}
  \langle X_s\rangle(r)
  = \bigl\langle X_s(t) \bigr\rangle_{\,r(t)\,\in\,[r,\,r+\mathrm{d}r)} .
  \label{eq:condavg}
\end{equation}

\begin{figure}[t]
  \centering
  \includegraphics[width=\columnwidth]{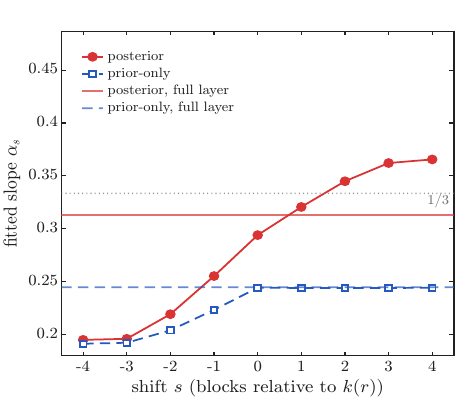}
  \caption{Fitted power-law exponent $\alpha_s$ of the conditioned time-integrated amplitude difference, $\langle X_s\rangle(r)\propto r^{\alpha_s}$, versus the layer shift $s$ relative to the separation-dependent cutoff $k(r)$. Red filled circles denote the posterior results, and blue open squares denote the prior-only results. The corresponding horizontal red solid and blue dashed lines indicate the posterior and prior-only all-layer reference values, respectively. The gray dotted line marks $\alpha=1/3$.}
  \label{fig:shift}
\end{figure}

If the MSM model's hidden layers correspond to physical eddies of a corresponding length scale, three things will happen. (i) Retaining all layers must reproduce the inertial-range scaling. (ii) If we randomize the inferred states of all layers finer than $k(r)+s$, retaining only the coarser layers, consistent with the Richardson picture, the inertial-range scaling is preserved for $s\geq 0$ and degrades rapidly once $s<0$. (iii) Repeating this randomization on prior-sampled states should reveal no $r^{1/3}$ scaling.


Fig.~\ref{fig:shift} shows the fitted exponent $\alpha_s$ in Eq. \ref{eq:r13} as a function of the shift $s$.  Three features stand out. First, the posterior all-layer reference, shown by the horizontal red solid line, gives $\alpha_A=0.313$, in close agreement with the Kolmogorov $1/3$ scaling. The full posterior reproduces the inertial-range scaling. Second, for $s<0$, the exponent collapses rapidly ($\alpha_{-2} = 0.219$). Randomizing blocks coarser than the separation decorrelates blocks that carry the $r$-dependent signal. In addition, at the physical truncation $s=0$, the exponent $\alpha_0 = 0.294$ has already recovered most of the scaling. This small residual reflects the non-locality of pair dispersion: the relative velocity at separation $r$ is set not sharply by eddies of scale $\sim r$ alone but by a band of scales centered on $r$ and extending to finer scales, so the hard cut at $k(r)$ strips the intermittency multipliers of these finer layers and discards their near-$r$ contribution, weakening the multiplicative fluctuations and holding the exponent slightly below the saturation value. Retaining additional fine-scale blocks increases the exponent above the all-layer value, reaching $\alpha_{+4}=0.366$.
Notably, the sharp asymmetry about the physically prescribed cutoff---scaling lost below it, only a slow residual gain above it---demonstrates that the posterior states are not merely a statistical re-description of the increment series: they carry quantitative spatial information about the cascade, block by block. 
Third, the prior-only null test fails to reproduce the Kolmogorov scaling. Its all-layer reference is only $\alpha_A^{\mathrm{prior}}=0.245$, well below $1/3$, which shows that the observed asymmetry in the posterior case does not arise by construction in the MSM model.

Our results demonstrate that the posterior hidden layer information extracted from a single trajectory does reveal spatial information about the ``turbulent eddies'' of different sizes around the trajectory. 
The prior-only null test confirms that the asymmetry under layer randomization originates from the hidden layer information recovered by the posterior. Taken together, these results show that the temporally inferred layer maps onto the physical hierarchy of ``eddy" scales, i.e., the layer associated with a given temporal scale corresponds to the spatial scale at which it contributes to pair separation. This temporal-to-spatial correspondence is what allows truncation at the separation-matched layer to recover the Kolmogorov scaling.



\emph{Discussion.}---We have demonstrated an MSM framework that infers hidden multiscale cascade states directly from a single Lagrangian trajectory and reproduces canonical intermittency statistics. Crucially, these inferred states encode not only temporal intermittency but also the spatial organization of the cascade. Selective randomization of the hidden layers reveals a scale-dependent response consistent with the Richardson--Obukhov picture of pair dispersion. A trajectory that is observed only in time, therefore, retains information about how the surrounding turbulent cascade is organized in space. This answers the central question posed here: despite containing no explicit information about the surrounding velocity field, a single Lagrangian trajectory can encode the spatial hierarchy of the cascade, with the temporally inferred layers corresponding to physical eddies at their associated length scales. This opens a route to spatial cascade diagnostics from purely Lagrangian observations, including particle-tracking experiments, atmospheric balloons, and oceanographic drifters for which an instantaneous Eulerian velocity field is unavailable.

\begin{acknowledgments}
This work was supported by the U.S. National Science Foundation CBET-2552483.
\end{acknowledgments}

\bibliographystyle{apsrev4-2}
\bibliography{references}

\end{document}